\documentclass[fleqn,twoside,twocolumn,nofootinbib,showkeys]{revtex4} 
\usepackage[sec,nocpr]{ujp} 

\begin{document}
\title[Self-Organization and Fractality in the Metabolic Process of
Glycolysis]
{SELF-ORGANIZATION AND FRACTALITY\\ IN THE METABOLIC PROCESS OF
GLYCOLYSIS}%
\author{V.I. Grytsay}
\affiliation{\bitp}
\address{\bitpaddr}
\email{vgrytsay@bitp.kiev.ua}

\udk{577.3} \pacs{05.45.-a, 05.45.Pq,\\[-3pt]
05.65.+b} \razd{\secx}

\autorcol{V.I.\hspace*{0.7mm}Grytsay}

\setcounter{page}{1}%

\begin{abstract}
Within a mathematical model, the metabolic process of glycolysis is
studied. The general scheme of glycolysis is considered as a natural
result of the biochemical evolution. By using the theory of
dissipative structures, the conditions of self-organization of the
given process are sought. The autocatalytic processes resulting in
the conservation of cyclicity in the dynamics of the process are
determined. The conditions of breaking of the synchronization of the
process, increase in the multiplicity of a cyclicity, and appearance
of chaotic modes are studied. The phase-parametric diagrams of a
cascade of bifurcations, which characterize the transition to
chaotic modes according to the Feigenbaum scenario and the
intermittence, are constructed. The strange attractors formed as a
result of the funnel effect are found. The complete spectra of
Lyapunov indices and divergences for the obtained modes are
calculated. The values of KS-entropy, horizons of predictability,
and Lyapunov dimensions of strange attractors are determined. Some
conclusions concerning the structural-functional connections in
glycolysis and their influence on the stability of the metabolic
process in a cell are presented.
\end{abstract}


\maketitle

\noindent Glycolysis is one of the oldest systems of biochemical
reactions that split hexoses. Possibly, it was formed in
protobionts, which were primary cells of the Earth (3.5--4~bln years
ago). Those cells were anaerobic heterotrophs. As a nutriment, they
used organic substances of the abiogenic origin, which were created
by chemo- and phototrophs. Moreover, they got the energy for
themselves from the oxidation and the fermentation in Earth's
primary oxygenless atmosphere. At the present time, the given
biochemical process as a result of the evolution is present
practically in all cells, which indicates its relict origin. Namely
from glycolysis starts the metabolic process of anaerobic catabolism
of glucose, which is completed by the formation of pyruvate. Then
the product of glycolysis can be used in three ways: the complete
oxidation to CO$_2 $ and water under aerobic conditions and the
fermentation to lactate or ethanol under anaerobic ones. Glycolysis
includes 10 successive reactions, which are running under the action
of enzymes in the cytoplasm of cells and are not connected with
membrane \mbox{systems.}

But the following questions permanently arise: How was the unique
stable sequence of the reactions of glycolysis running in cytoplasm,
amorphous at first glance, formed from the huge number of organic
substances present in the primary broth and which mechanism of their
selection was? In author's opinion, the answers should be sought on
the basis of the general theory of chemical evolution by Professor
A.P.~Rudenko \cite{1,2,3}.

This theory allows one to solve the problems related to the moving
forces and the mechanism of evolution in catalytic systems. On this
way, the laws of the chemical evolution and the selection of
elements and structures, as well as their causal dependence, are
studied. The complexity of a chemical organization and a hierarchy
of chemical systems are a consequence of the evolution. With
reagents, catalysts form an intermediate (multiplet) complex
possessing the properties of a transient state. Such complex exists
in the form of continuously varying configurations on some small
section of the reaction path. However, in the course of the
catalytic reactions accompanied by a constant inflow of new portions
of reagents and an outflow of ready products, the complexes are
multiply reproduced. They take the status of elementary open
catalytic systems (EOCSs). Rudenko indicated the particular
dynamical type of stability of such systems. It can be
quantitatively characterized in terms of the intensity of exchange
of substances and the energy of the basic reaction, which is equal
to the product of the activity of a catalytic center by the
elementary chemical affinity of the basic
\mbox{reaction.}\looseness=1

Thus, Rudenko made conclusion that there occurs the natural selection of the
catalytic centers with the highest activity in the process of
self-development of EOCSs. On those centers, the basic reaction is
concentrated more and more. The centers with lower activity are gradually
eliminated from the kinetic continuum and ``do not survive''. At the
multiple successive changes in EOCSs, the transition to a higher level of
steadiness is accompanied by the evolution of the mechanism of the basic
reaction due to changes in the composition and the structure of a catalyst
operating in the beginning of the reaction, as well as due to the division
of the chamical process into elementary stages and the appearance of new
catalysts of these stages due to changes in EOCSs.

The above consideration allows us to make the following conclusion.
The validity of Rudenko's theory of chemical evolution of elementary
open catalytic systems is supported by its correspondence with the
general theory of dissipative structures \cite{4}. Just the
dissipative structures arising in open nonlinear systems of the
Nature are the reason for the self-organization in it. At the
appearance of an autocatalytic oscillatory process in such
dissipative system, the system becomes self-developing. Those paths
of evolutionary changes are formed with the highest rate and the
probability, on which the absolute catalytic activity increases
maximally.

During the chemical evolution on the Earth at the appearance of favorable
conditions for the creation of organic substances, the enzymatic reactions
replaced rapidly and completely the inorganic catalysis. In the primary
broth, the almost infinite number of open biochemical systems were formed.
But only those of them were evolved, in which the autocatalysis was most
intense. As a result of the self-organization of the given elementary
biochemical processes, the metabolic networks with the ability to a
self-assembling were formed in larger interconnected open nonlinear systems.
This occurred until a cell was created. One of the biochemical processes
conserved from the previous metabolism is glycolysis.

A lot of experimental and theoretical works are devoted to the study
of glycolysis. The researchers were expecially interested in the
glycolytic oscillations observed in yeast cells \cite{5}. The
experimental characteristics were obtained, and a number of
mathematical models of this process were constructed \cite{6}.
Sel'kov proposed a fine theoretical model \cite{7}, in which the
enzyme phosphofructokinase is activated by its. Then Goldbeter and
Lefever developed a more detailed model \cite{8}, in which the
allosteric nature of phosphofructokinase was taken into account.

The other mathematical models of biochemical processes can be found,
for example, in \cite{9,10,10a,10b,10c,11}. The obtained results
enrich our knowledge about the given processes.

The study presented in this work will be based on the mathematical
model of glycolysis and gluconeogenesis constructed by Professor
V.P.~Gachok \cite{11a,12,13}. A specific feature of his model
consists in the consideration of the action of such factors as the
adeninenucleotide cycle and the feedback of gluconeogenesis on the
allosteric enzyme phosphofructokinase. This allowed him to analyze
more qualitatively the reasons for the appearance of an oscillatory
dynamics in glycolysis.

\begin{figure*}%
\vskip1mm
\includegraphics[width=12.0cm]{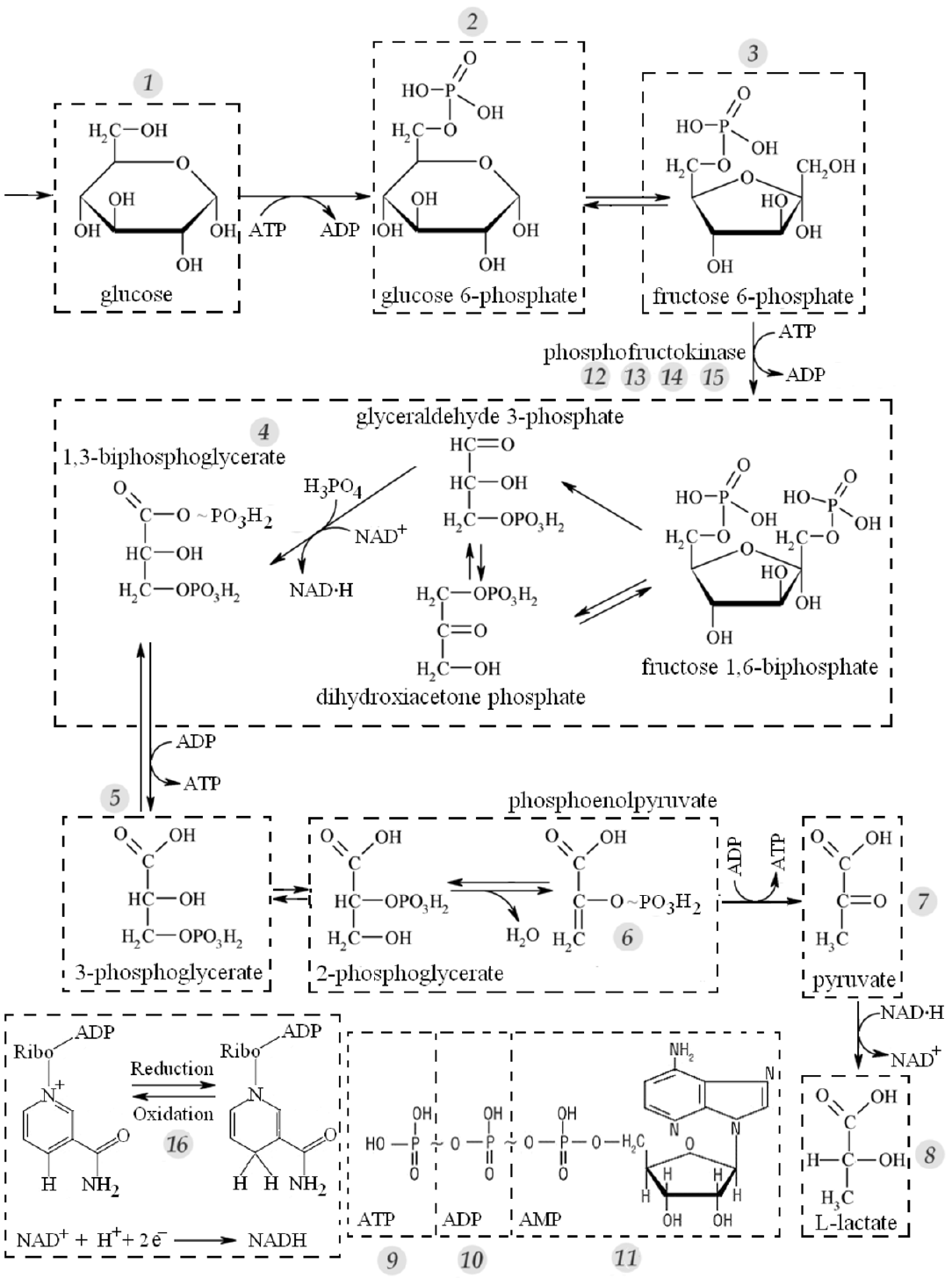}
\vskip-3mm\parbox{12.0cm}{\caption{General scheme of the metabolic
process of glycolysis\label{fig:1}}}\vspace*{0.5mm}
\end{figure*}

In what follows, the model is improved so that the complete chain of
the whole metabolic process of glycolysis from start to end under
anaerobic conditions is modeled. This enables us to study the
positive feedbacks of the process, which create a stable
autocatalytic process of the given section of metabolism
irrespective of other metabolic processes in a cell. Glycolysis is
considered as an open part of the biosystem, which is self-organized
by itself due to the input substances and output products of the
reaction in a cell, ensuring the condition of its survival and
\mbox{evolution.}\looseness=1\vspace*{-2mm}

\section{Mathematical Model}\vspace*{-0.5mm}

The general scheme of the process of glycolysis is presented in
Fig.~\ref{fig:1}. According to it, the mathematical model
(\ref{eq1})--(\ref{eq16}) is constructed with regard for the mass
balance and the enzymatic kinetics.

Here, we will describe the process of catabolism of glucose to
lactate under anaerobic conditions. The insignificant changes in
Eq.\,(\ref{eq8}) and the replacement of the enzyme lactate
dehydrogenase by pyruvate decarboxylase allow the application of
this model to the study of the process of alcoholic fermentation. In
this case, the model will describe the formation of ethanol instead
of lactate. The internal dynamics of process given by the solution
of Eq.\,(\ref{eq1})--(\ref{eq16}) remains \mbox{invariable}.
\begin{equation}
\label{eq1}
\frac{dG}{dt}=\frac{G_0 }{S}\frac{m_1 }{m_1 +F_1 }-l_8 V(G)V(T),
\end{equation}\vspace*{-7mm}
\[
\frac{dF_1 }{dt}=l_8 V(G)V(T)-l_1 V(R_{_1 } )V(F_1 )V(T)\,+
\]\vspace*{-7mm}
\begin{equation}
\label{eq2} +\,l_5 \frac{1}{1+\gamma A}V(F_2 )-m_3 \frac{F_1 }{S},
\end{equation}\vspace*{-7mm}
\begin{equation}
\label{eq3}
\frac{dF_2 }{dt}=l_1 V(R_1 )V(F_1 )V(T)-l_5 \frac{1}{1+\gamma A}V(F_2 )-m_5
\frac{F_2 }{S},
\end{equation}\vspace*{-7mm}
\[
\frac{d\psi _1 }{dt}=\frac{m_5 (F_2 /S)}{S_1 +m_5 (F_2 /S)}-l_6
V(\psi _1 )V(D)\,+
\]\vspace*{-7mm}
\begin{equation}
\label{eq4} +\,m_7 V(M-N)V(P),
\end{equation}\vspace*{-7mm}
\begin{equation}
\label{eq5}
\frac{d\psi _2 }{dt}=l_6 V(\psi _1 )V(D)-m_8 \frac{\psi _2 }{S},
\end{equation}\vspace*{-7mm}
\begin{equation}
\label{eq6}
\frac{d\psi _3 }{dt}=\frac{\psi _2 }{S}\frac{m_2 }{m_2 +\psi _3 }-l_2 V(\psi
_3 )V(D)-m_4 \frac{\psi _3 }{S},
\end{equation}\vspace*{-7mm}
\begin{equation}
\label{eq7}
\frac{dP}{dt}=l_2 V(\psi _3 )V(D)-m_6 \frac{P}{S}-l_7 V(N)V(P),
\end{equation}\vspace*{-7mm}
\begin{equation}
\label{eq8}
\frac{dL}{dt}=l_7 V(N)V(P)-m_9 \frac{L}{S},
\end{equation}\vspace*{-7mm}
\[
\frac{dT}{dt}=l_2 V(\psi _3 )V(D)-l_1 V(R_1 )V(F_1 )V(T)\,+
\]\vspace*{-7mm}
\[
+\,l_3 \frac{A}{\delta +A}V(T)-l_4 \frac{T^4}{\beta +T^4}\,+
\]\vspace*{-7mm}
\begin{equation}
\label{eq9} +\,l_6 V(\psi _1 )V(D)-l_9 V(G)V(T),
\end{equation}\vspace*{-7mm}
\[
\frac{dD}{dt}=l_1 V(R_1 )V(F_1 )V(T)-l_2 V(\psi _3 )V(D)\,+
\]\vspace*{-7mm}
\[
+\,2^\ast l_3 \frac{A}{\delta +A}V(T)-l_6 V(\psi _1 )V(D)\,+
\]\vspace*{-7mm}
\begin{equation}
\label{eq10} +\,l_9 V(G)V(T),
\end{equation}\vspace*{-7mm}
\begin{equation}
\label{eq11}
\frac{dA}{dt}=l_4 \frac{T^4}{\beta +T^4}-l_3 \frac{A}{\delta +A}V(T),
\end{equation}\vspace*{-7mm}
\[
\frac{dR_1 }{dt}=k_1 T_1 V(F_1^2 )+k_3 R_2 V(D^2)\,-
\]\vspace*{-7mm}
\begin{equation}
\label{eq12} -\,k_5 R_1 \frac{T}{1+T+\alpha A}-k_7 R_1 V(T^2),
\end{equation}\vspace*{-7mm}
\[
\frac{dR_2 }{dt}=k_5 R_1 \frac{T}{1+T+\alpha A}-k_3 R_2 V(D^2)\,+
\]\vspace*{-7mm}
\begin{equation}
\label{eq13} +\,k_2 T_2 V(F_1^2 )-k_8 R_2 V(T^2),
\end{equation}\vspace*{-7mm}
\[
\frac{dT_1 }{dt}=k_7 R_1 V(T^2)-k_6 T_1 \frac{T}{1+T+\alpha A}\,+
\]\vspace*{-7mm}
\begin{equation}
\label{eq14} +\,k_4 T_2 V(D^2)-k_1 T_1 V(F_1^2 ),
\end{equation}\vspace*{-7mm}
\[
\frac{dT_2 }{dt}=k_6 T_1 \frac{T}{1+T+\alpha A}-k_4 T_2 V(D^2)\,-
\]\vspace*{-7mm}
\begin{equation}
\label{eq15} -\,k_2 T_2 V(F_1^2 )+k_8 R_2 V(T^2),
\end{equation}\vspace*{-7mm}
\begin{equation}
\label{eq16}
\frac{dN}{dt}=-l_7 V(N)V(P)+l_7 V(M-N)V(\psi _1 ).
\end{equation}
Here, $V(X)=X/(1+X)$ is the function accounting for the adsorption
of the enzyme in the region of local coupling. The variables of the
system of equations are made dimensionless \cite{11a,12,13}.

We take the following values of parameters:
\[
l_1 =0.0535; \quad l_2 =0.046; \quad l_3 =0.0017;
\]\vspace*{-9mm}
\[
 l_4
=0.01334; \quad l_5 =0.3; \quad l_6 =0.001;\quad l_7 =0.01;
\]\vspace*{-9mm}
\[
 l_8 =0.0535; \quad l_9 =0.001; \quad k_1 =0.07; \quad k_2
=0.01;
\]\vspace*{-9mm}
\[
  k_3 =0.0015;
\quad
 k_4 =0.0005; \quad k_5 =0.05;
\]\vspace*{-9mm}
\[
k_6 =0.005;\quad k_7 =0.03; \quad k_8 =0.005;  \quad m_1 =0.3;
\]\vspace*{-9mm}
\[
m_2 =0.15; \quad m_3 =1.6; \quad m_4 =0.0005;\quad m_5 =0.007;
\]\vspace*{-9mm}
\[
m_6 =10; \quad m_7 =0.0001; \quad m_8 =0.0000171;
\]\vspace*{-9mm}
\[
m_9 =0.5; \quad G_0 =18.4; \quad L=0.005;\quad S=1000;
\]\vspace*{-9mm}
\[
A=0.6779; \quad M=0.005; \quad S_1 =150; \quad \alpha =184.5;
\]\vspace*{-9mm}
\[
\beta =250; \quad \delta =0.3; \quad \gamma =79.7.
\]
The mathematical model is given by the system of nonlinear
differential equations. The equations correspond to basic sections
of the metabolic process, which define the sequence of reactions and
affect the stability of glycolysis. Some parts of the metabolic
network, which affect insignificantly the self-organization of the
process, are described generically. In Fig.~\ref{fig:1}, we show the
parts of the metabolic network (from 1 to 16) corresponding to the
number of a differential equation: (\ref{eq1})--(\ref{eq16}).

On the first stage (\ref{eq1}), the incoming substance $G_0 $
(glucose) is phosphorylated with the help of the enzyme hexokinase
to glucose-6-phosphate. The donor of a phosphorylic group is a
molecule ATP $(T)$ (\ref{eq1}), (\ref{eq9}). This reaction is
running irreversibly. The mo\-le\-cu\-les of
glu\-co\-se-6-phos\-phate cannot leave a cell. In this case, they
are a product of the reaction and an allosteric inhibitor. When the
concentration of glu\-co\-se-6-phos\-phate in the cell exceeds the
normal level, glu\-co\-se-6-phos\-phate inhibits temporarily and
reversibly hexokinase (\ref{eq1}) $(F_1 )$, so that the rate of its
formation is put in correspondence with the rate of its consumption
in the subsequent reaction. Then, with the participation of the
enzymes phosphohexoisomerase (or phosphoglucoseisomerase), the
reversible isomerization of glucose-6-phosphate in
fructose-6-phosphate, which does not influence the irreversibility
of the process, occurs.

Equations (\ref{eq2}), (\ref{eq3}) describe the process of formation
of fructose-6-phosphate ($F_1 $) and its transformation into
fructose-1,6-diphosphate ($F_2 $). This happens under the catalytic
action of the key enzyme of glycolysis~-- phosphofructokinase. This
enzyme catalyzes the irreversible transfer of a phosphorylic group
from ATP (\ref{eq2}), (\ref{eq9}) to fructose-6-phosphate with the
formation of fructose-1,6-diphosphate. The substrate
fructose-6-phosphate is an activator, whereas ATP is an inhibitor of
the given process. In addition to such regulation, the given enzyme
can be regulated by the adeninenucleotidic cycle: ATP--ADP--AMP (see
below), which favors the support of an optimal stable stationary
state. We now describe the process of gluconeogenesis. At low
concentrations of reagents, the reaction is reversible $F_2 \to F_1
$ (\ref{eq2}), (\ref{eq3}), and the positive feedback affecting the
stability of the process is created. The subsequent splitting of
fructose-1,6-diphosphate into two different triosephosphates
(glyceraldehyde-3-phosphate and dihydroxyacetonephosphate) occurs
\mbox{reversibly.}

Equation (\ref{eq4}) describes the formation of
1,3-di\-phos\-pho\-gly\-ce\-ra\-te ($\psi _1 )$, which is the start
of the second stage of glycolysis. With the help of the enzyme
$D$-gly\-ceraldehydephosphate dehydrogenase,
glyceraldehyde-3-phosphate is oxidized and joins phosphoric acid. In
this case, the role of an acceptor of hydrogen is played by coenzyme
$\textrm{NAD}^+$. There occurs the enzymatic reduction:
$\textrm{NAD}^+\to \textrm{NAD}\cdot \textrm{H}$ (\ref{eq4}),
(\ref{eq16}).

With the help of Eq.\,(\ref{eq5}), we describe the process of
transfer of a high-energy phosphorylic group by the enzyme
phosphoglycerate kinase from a carboxyl of 1,3-diphosphoglycerate
onto ADP. As a result, ATP (\ref{eq11}) and 3-phosphoglycerate $\psi
_2 $ are formed.

Equation (\ref{eq6}) describes the formation of
2-phos\-pho\-gly\-ce\-ra\-te with the help of the enzyme
phos\-pho\-gly\-ce\-ra\-te mutase. Then water is eliminated, which
results in the formation of phos\-pho\-enol\-py\-ru\-va\-te $\psi _3
$.

The formation of pyruvate $P$ under the action of the enzyme
pyruvate kinase is presented by Eq.\,(\ref{eq7}). Here, the
phosphorylation of the substrate occurs. As distinct from reactions
(\ref{eq4})--(\ref{eq5}), where the reverse reaction of
gluconeogenesis at the synthetic absorption of CO$_2 $ is possible,
we observe the powerful irreversible process: $\textrm{ADP}\to
\textrm{ATP}$.

Equation (\ref{eq8}) characterizes the formation of thr second
product of the process~-- lactate $L$. With the help of the enzyme
lactate dehydrogenase, there occurs the enzymatic oxidation
$\textrm{NAD}\cdot \textrm{H}\to \textrm{NAD}^+$. The balance
between $\textrm{NAD}^+$ and $\textrm{NAD}\cdot \textrm{H}$ holds
(\ref{eq16}).

Equations (\ref{eq9})--(\ref{eq11}) describe the kinetics of changes
in the contents of ATP (\ref{eq9}), ADP (\ref{eq10}), and AMP
(\ref{eq11}) according to the metabolic scheme of glycolysis (see
above). On the whole, the adeninenucleotidic cycle of mutual
transitions between the given reagents arises: ATP--ADP--AMP. This
cycle favors the conservation of the optimum stationary state of the
metabolic \mbox{process.}

Equations (\ref{eq12})--(\ref{eq15}) show the kinetics of changes in
the content of the allosteric enzyme phosphofructokinase. It is
assumed that the enzyme has two active forms ($R_1 $ (\ref{eq12})
and $R_2 $ (\ref{eq13})) and two inactive ones ($T_1 $ (\ref{eq14})
and $T_2 $ (\ref{eq15})). In this case, we observe the mutual
transformation of forms $T_1 $ and $R_1 $, $T_2 $ and $R_2 $. The
equations present the general scheme of regulating connections. Form
$R_1 $ (\ref{eq12}) is created from form $T_1 $ as a result of the
saturation of two allosteric centers by molecules $F_1 $ and from
form $R_2 $ at the expense of two molecules $D$. The inactivation of
form $R_1 $ occurs at the expense of $T$(\ref{eq12}) with the
formation of form $R_2 $ (\ref{eq13}) and two molecules
$T$(\ref{eq12}) with the formation of form $T_2 $ (\ref{eq15}). This
invertible inactivation is inhibited with increase in $A$ according
to the high level of $T$ (parameter $\alpha )$ (\ref{eq12}).
Equations (\ref{eq13})--(\ref{eq15}) are constructed
ana\-logously.\looseness=1

Equation (\ref{eq16}) describes the kinetics of changes in the
content of the reduced form nicotineamideadeninenucleotide
$\textrm{NAD}\cdot \textrm{H}$, according to its consumption and the
reduction of oxidized form $\textrm{NAD}^+$ (\ref{eq4}). The balance
between the oxidized and reduced forms in the glycolytic cycle is
conserved in the invariable form. In this case, the integral of
motion $\textrm{NAD}\cdot \textrm{H}(t)+\textrm{NAD}^+(t)=M$ is
satisfied.

\begin{figure*}%
\vskip1mm
\includegraphics[width=16.7cm]{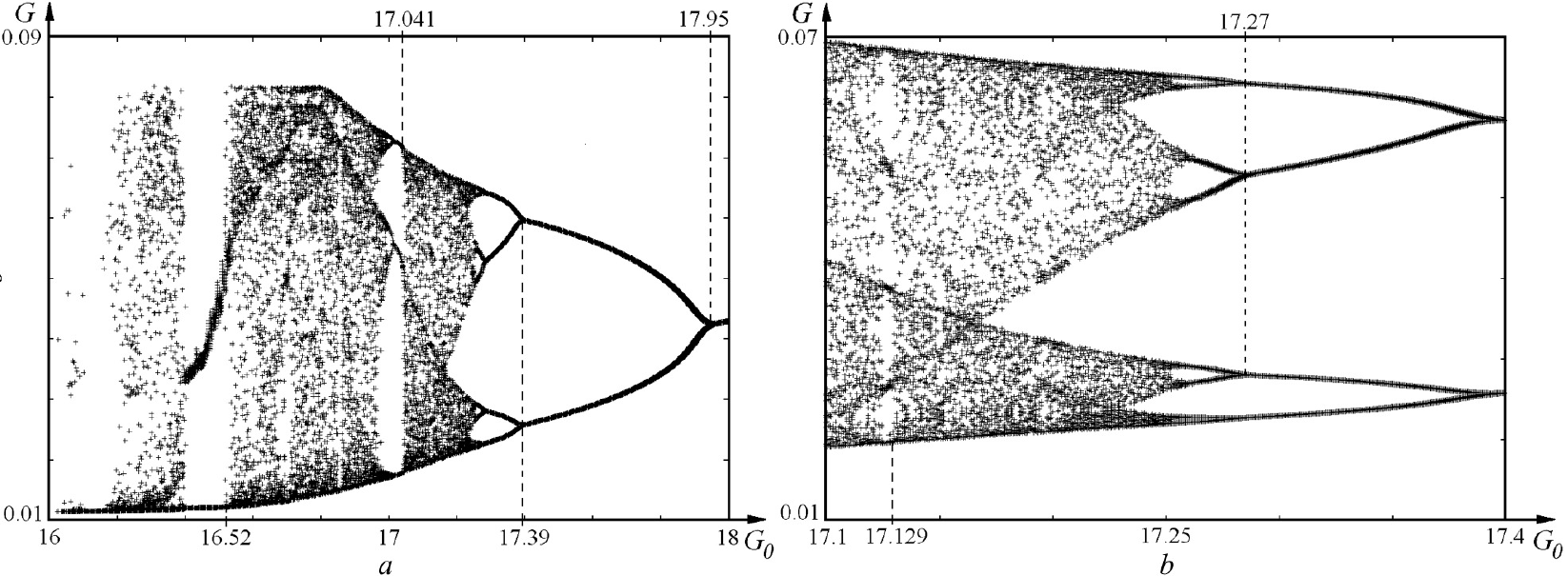}
\vskip-3mm\caption{Phase-parametric diagram of the system for the
variable $G(t)$: \textit{a}~-- $G_0 \in (16, 18)$; \textit{b}~--
$G_0 \in (17.1, 17.4)$}\label{fig:2}\vspace*{-1.5mm}
\end{figure*}

The study of solutions of the present mathematical model
(\ref{eq1})--(\ref{eq16}) is realized with the help of the theory of
nonlinear differential equations \cite{14,15} and the methods of
modeling of biochemical systems used earlier in works
\cite{16,17,18,19,20,21,22,23,24,25,26,27,28,29,30,31,32,33,34,35,36}.

The numerical solution of this autonomous system of nonlinear differential
equations wasmade within the Runge--Kutta--Merson method. The accuracy of
calculations was set to be $10^{-8}$. For the reliability of studies, namely
for the system being on the transient initial state to approach the
asymptotic solution with an attractor, the duration of calculations was
taken to be $10^6$. For this time, the trajectory ``sticks'' the appropriate
attractor.

\section{Results of Studies}

The given mathematical model is a system of nonlinear differential
equations (\ref{eq1})--(\ref{eq16}), which describes the open
nonlinear biochemical system. For it, the input substance is glucose
characterized by the coefficient $G_0 $. The output products of the
reaction are lactose, ATP, and $\textrm{H}_2 \textrm{O}$. Namely the
flows of these substances form the internal dynamics of the given
metabolic process. At the breaking of the mass balance between them,
the continuity of the running of glycolysis is violated as well.
From the energetic viewpoint, the transformation of glucose in
pyruvate means a significant decrease in the free energy of the
products of the reaction. Therefore, glycolysis is the energy-gained
irreversible process running in the open nonlinear system far from
equilibrium. In addition, the whole metabolic process of glycolysis
is embraced by a feedback formed by the redox reactions of transfer
of electrons with the help of NAD (\ref{eq16}) (Fig.~\ref{fig:1}).
Due to NAD, glycolysis includes the autocatalytic process of
catabolism of glucose. After the successive splitting of a molecule
of glucose and the appearance of 2 molecules (ATP and a molecule of
pyruvate) on the output, the system returns in the initial state.
Thus, the metabolic process of glycolysis in a cell can be separated
as a united self-regulating complex. The whole metabolic process of
glycolysis can be considerd as the process of self-organization,
which is functioning in the cyclic mode. When the evolution of
metabolic processes in protobionts was completed, glycolysis remains
to be invariable in all~cells.

\begin{figure*}%
\vskip1mm
\includegraphics[width=16.7cm]{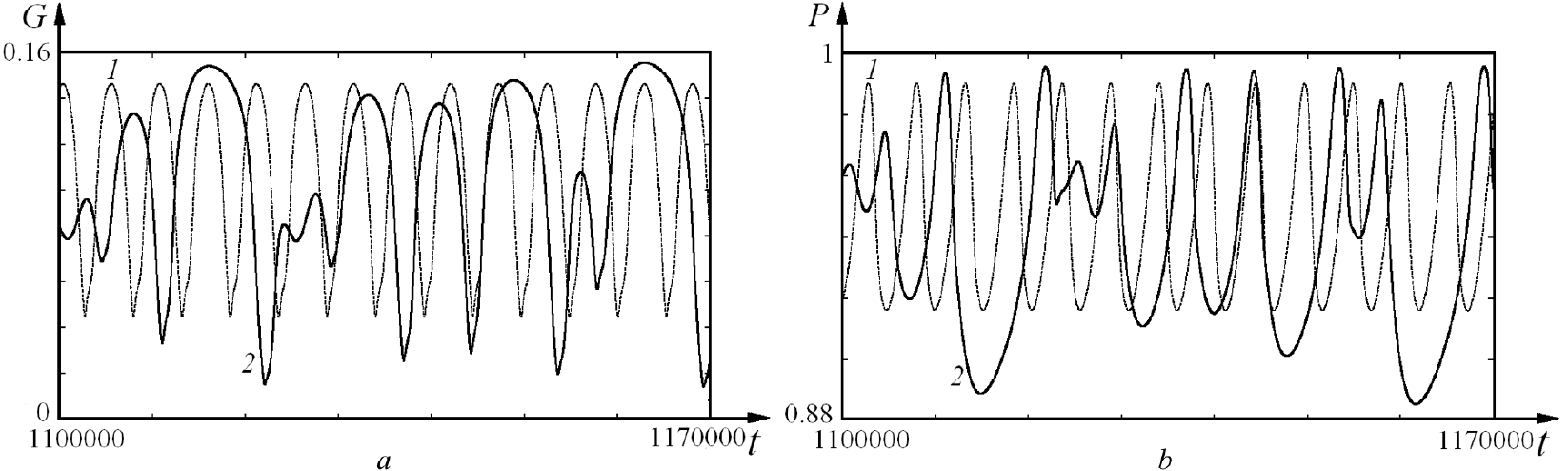}\\[2mm]
\includegraphics[width=16.7cm]{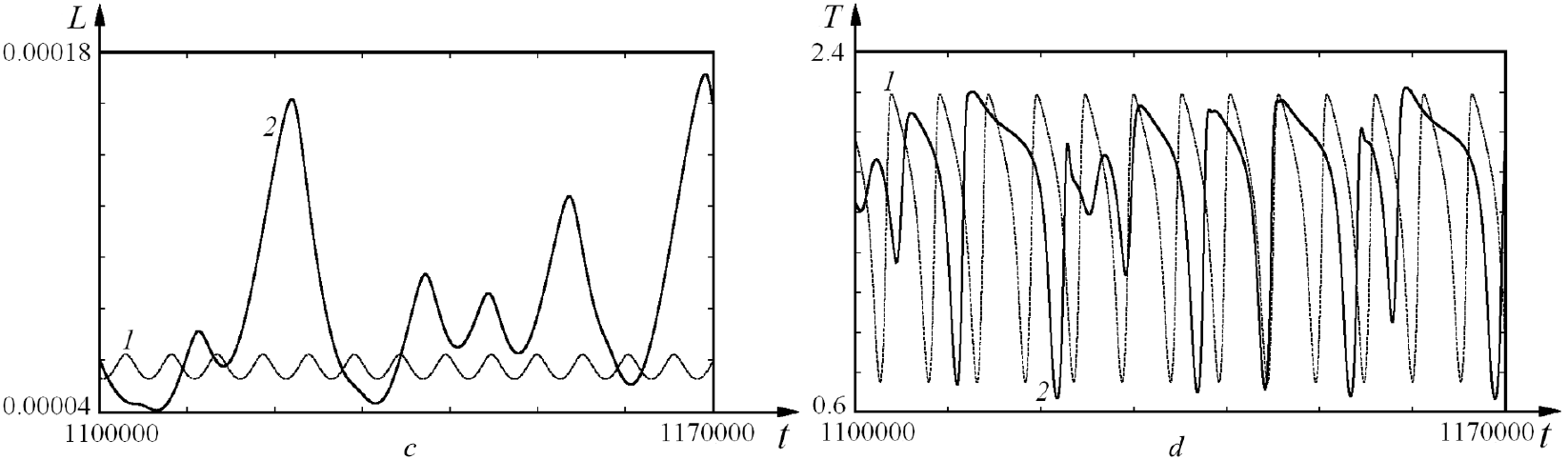}
\vskip-3mm\caption{Kinetic curves for the variables: $G(t)$
(\textit{a}), $P(t)$ (\textit{b}), $L(t)$ (\textit{c}), and $T(t)$
(\textit{d}) in the 1-fold periodic mode for $G_0 =19$ (\ref{eq1})
and in the chaotic mode for $G_0 =16.7$
(\ref{eq2})}\label{fig:3}\vspace*{-2mm}
\end{figure*}

Let us study the dependence of the oscillatory dynamics of the
metabolic process of glycolysis on $G_0 $. Figure
\ref{fig:2},~\textit{a},~\textit{b} presents the constructed
phase-parametric diagrams of the system for the variable $G(t)$,
when $G_0 $ changes in the corresponding intervals. To construct the
phase-parametric diagrams, we used the method of cuts. In the phase
space containing the trajectories of the system, we place a cutting
plane at $R_1 =0.7$. If the trajectory crosses this plane in a
certain direction, the value of the chosen variable (in this case,
$G(t)$) is placed on the phase-parametric diagram. Such choice is
explained by the symmetry of oscillations of the active form of the
allosteric enzyme phosphofructokinase relative to this point in many
modes calculated earlier. For every given value of $G_0 $, we mark
the intersection of this plane in a single direction by the
trajectory, when the trajectory has approached the attractor. In the
case where a multiple periodic limiting cycle arises, we mark a
number of points on the plane. These points coincide in the period.
If a deterministic chaos arises, then such points, where the
trajectory intersects the plane, are positioned chaotically.

Consider the diagrams from right to left. It is seen from phase-parametric
diagrams that, for $G_0^j =17.95$, the period of oscillations is doubled.
For $G_0^{j+1} =17.39$, we observe the repeated doubling of the period.
Then, for $G_0^{j+2} =17.27$, the period of autooscillations is doubled once
more. As $G_0 $ decreases further, no doubling of the period of
autooscillations occurs, and a chaotic mode arises as a result of the
intermittence. The determined sequence of bifurcations satisfies the
relation
\[
\mathop {\lim }\limits_{t\to \infty } \frac{G_0^{j+1} -G_0^j }{G_0^{j+2}
-G_0^{j+1} }\approx 4.667.
\]
This number is very close to the universal Feigenbaum constant.
Thus, as the parameter $G_0 $ decreases on this section, the
doubling of the period of autooscillations occurs according to the
Feigenbaum scenario \cite{37}. This means that, in the given
unstable modes of the physical system, any arising fluctuation can
induce a chaotic cyclic mode.

For $G_0 =17.129$ (Fig.~\ref{fig:2},~\textit{b}) and $G_0 =17.041$,
$G_0 =16.52$ (Fig.~\ref{fig:2},~\textit{a}), the windows of
periodicity arise. The deterministic chaos is destroyed, and
periodic and quasiperiodic modes are established. Outside these
windows, chaotic modes arise. The identical windows of periodicity
are observed also on less scales of the diagram. In other words, the
phase-parametric diagrams on small and large scales are analogous.
This indicates the fractal nature of the obtained cascade of
bifurcations in the metabolic process of glycolysis.

\begin{figure*}%
\vskip1mm
\includegraphics[width=16.7cm]{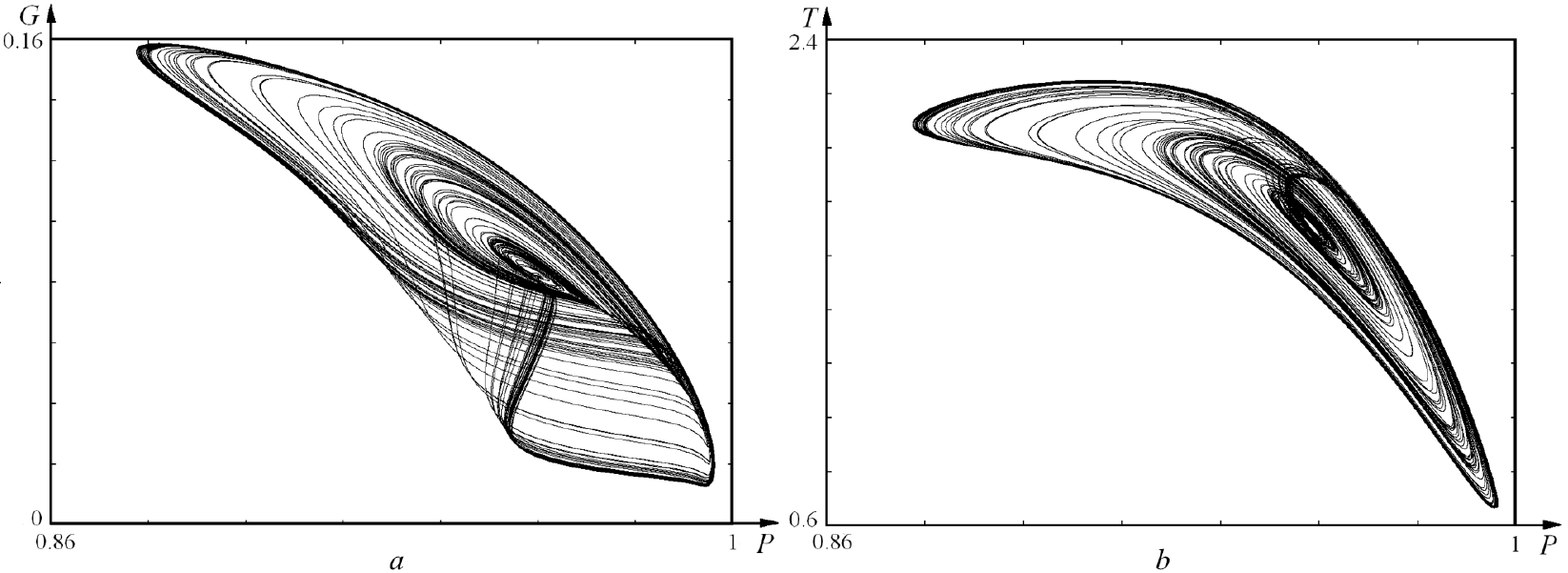}\\[2mm]
\includegraphics[width=16.7cm]{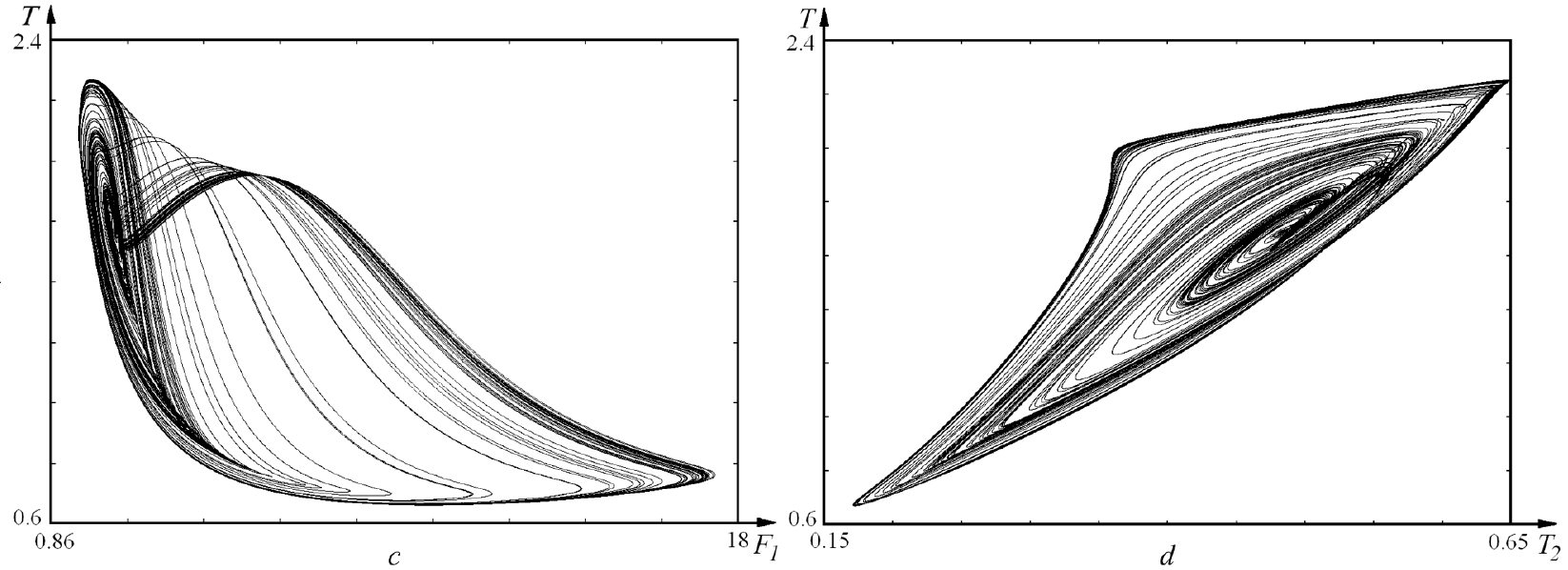}
\vskip-3mm\caption{Projections of the phase portrait of a strange
attractor formed for $G_0 =16.7$ in the appropriate plane: $(P,G)$
\textit{a}, $(P,T)$  \textit{b}, $(F_1 ,T)$  \textit{c}, and $(T_2
,T)$ \textit{d}}\label{fig:4}
\end{figure*}

In Fig.~\ref{fig:3},~\textit{a--d}, we show, as an example, the
kinetics of autooscillations for some components of the metabolic
process in the 1-fold periodic mode for $G_0 =19$ and in the chaotic
mode for $G_0 =16.7$. The synchronous autooscillations of glucose,
pyruvate, lactate, and $ATP$ become chaotic.

According to the kinetics, the phase portraits of the system are
changed as well. In Fig.~\ref{fig:4},~\textit{a--d}, we present, as
an example, some projections of phase portraits of the system for
$G_0 =16.7$. The given chaotic mode is a strange attractor formed as
a result of the funnel effect. Inside the formed funnel,, we observe
the mixing of dispersing and approaching one another trajectories.
Due to fluctuations, the stability of the cycle is violated, and the
cycle becomes chaotic. A deterministic chaos is formed.

While studying the phase-parametric diagrams in
Fig.~\ref{fig:2},~\textit{a},~\textit{b}, it is impossible
beforehand to determine, at which values of the parameter $G_0 $ a
multiple stable (quasistable) autoperiodic cycle or a strange
attractor is formed.

For the unique identification of the type of obtained attractors and
for the determination of their stability, we calculated the complete
spectra of Lyapunov indices and their sums $\Lambda =\sum
_{j=1}^{16} \lambda _j $ at some chosen points. The calculation was
carried out by Benettin's algorithm with the orthogonalization of
the perturbation vectors by the Gram--Schmidt
method~\cite{15}.\looseness=1

A specific feature of the calculation of those indicators consists
in the complexity of the determination of the perturbation vectors
represented by $16\times 16$ matrices on a personal computer.

The algorithm of calculation of the complete spectrum of Lyapunov
indices is as follows. First, some point on the attractor on the
attractor $\overline {X_0 } $ is taken as the input one. Then the
trajectory going out from the point and the evolution of $N$
perturbation vectors are traced. In our vase, $N = 16$ is the number
of variables of the system. The input equations of the system, which
are supplemented by 16 complexes of equations in variations, are
solved numerically. As the initial perturbation vectors, we set the
collection of vectors $\overline {b_1^0 } $, $\overline {b _2^0} $,
..., $\overline {b_{16}^0 } $, which are orthogonal to one another
and are normalized to 1. In some time $T$, the trajectory comes to
the point $\overline {X_1 } $, and the perturbation vectors are
$\overline {b_1^1 } $, $\overline {b_2^1 } $, ..., $\overline
{b_{16}^1 } $. Their renormalization and orthogonalization by the
Gram--Schmidt method are realized by the following scheme:
\[
\overline {b_1^1 } =\frac{\overline {b_1 } }{\left\| {\overline {b_1
} } \right\|},
\]\vspace*{-7mm}
\[
\overline {{b}'_2 } =\overline {b_2^0 } -(\overline {b_2^0 }
,\overline {b_1^1 } )\overline {b_1^1 } , \quad \overline {b_2^1 }
=\frac{\overline {{b}'_2 } }{\left\| {\overline {{b}'_2 } }
\right\|},
\]\vspace*{-7mm}
\[
\overline {{b}'_3 } =\overline {b_3^0 } -(\overline {b_3^0 }
,\overline {b_1^1 } )\overline {b_1^1 } -(\overline {b_3^0 }
,\overline {b_2^1 } )\overline {b_2^1 } , \quad \overline {b_3^1 }
=\frac{\overline {{b}'_3 } }{\left\| {{b}'_3 } \right\|},
\]\vspace*{-7mm}
\[
\overline {{b}'_4 } =\overline {b_4^0 } -(\overline {b_4^0 }
,\overline {b_1^1 } )\overline {b_1^1 } -(\overline {b_4^0 }
,\overline {b_2^1 } )\overline {b_2^1 } -(\overline {b_4 }
,\overline {b_3^1 } )b_3^1 , \quad \overline {b_4^1 }
=\frac{\overline {{b}'_4 } }{\left\| {{b}'_4 } \right\|},
\]
{...}{...}{...}{...}{...}{...}{...}{...}{...}{...}{...}{...}{...}{...}{...}{...}{...}{...}{...}{...}{...}{...}{...}{...}{...}{...}{...}\[
\overline {{b}'_{16} } =\overline {b_{16}^0 } -(\overline {b_{16}^0
} ,\overline {b_1^1 } )\overline {b_1^1 } -(\overline {b_{16}^0 }
,\overline {b_2^1 } )\overline {b_2^1 } -(\overline {b_{16} }
,\overline {b_3^1 } )b_3^1 -...-
\]\vspace*{-7mm}
\[
-\,(\overline {b_{16} } ,\overline {b_{15}^1 } )b_{15}^1 , \quad
\overline {b_{16}^1 } =\frac{\overline {{b}'_{16} } }{\left\|
{{b}'_{16} } \right\|},
\]
Then the calculation is continued, by starting from the point
$\overline {X_1 } $ and the perturbation vectors $\overline {b_1^1 }
$, $\overline {b_2^1 } $, ..., $\overline {b_{16}^1 } $. In the next
time interval $T$, a new collection of perturbation vectors
$\overline {b_1^2 } $, $\overline {b_2^2 } $, ..., $\overline
{b_{16}^2 } $ is formed. It is again orthogonalized and renormalized
by the above scheme. This sequence of operations is repeated a
sufficiently large number of times M. In this case, in the course of
calculations, we determine the sums
\[
S_1 =\sum\limits_{i=1}^M {\ln \left\| {{b}_1^{\prime i} } \right\|}
, \quad S_2 =\sum\limits_{i=1}^M {\ln \left\| {{b}_2^{\prime i} }
\right\|} ,~ ...,
\]\vspace*{-7mm}
\[
S_{16} =\sum\limits_{i=1}^M {\ln \left\| {{b}_{16}^{\prime i} }
\right\|} ,
\]
in which the perturbation vectors before the renormalization, but
after the orthogonalization are present.

We evaluate 16 Lyapunov indices in the following way:\vspace*{-3mm}
\[
\lambda _j =\frac{Z_j }{MT}, \quad j=1, 2, ..., 16.
\]
For comparison, we give the spectra of Lyapunov indices for some
modes of the system. For brevity without any loss of information,
the values of indices are rounded to the 5-th decimal point.

The criterion of validity of the calculation is the ratio of
Lyapunov indices: $\lambda _1 >\lambda _2 >\lambda _3 >...>\lambda
_{16} $. For a regular attractor, it must be: $\lambda _1 \approx
0$. The next subsequent indices can be also $\approx $0 in some
cases. In other cases, they are negative. The zero value of first
Lyapunov index indicate the presence of a stable limiting cycle.

For a strange attractor, the presence of at least one positive
Lyapunov index is obligatory. Then the zero index is positioned; the
rest indices are negative. The presence of negative indices means
the compression of the phase space in the corresponding directions,
whereas the positive indices indicate the divergence of trajectories
in some directions. Therefore, there occurs the mixing of
trajectories in narrow places of the phase space, and the
deterministic chaos arises. The Lyapunov indices include
obligatorily the zero index. This means the conservation of an
aperiodic trajectory of the attractor in some domain of the phase
space, which is the condition of existence of a strange attractor.

As an example for comparison, we now give several calculations of
the complete spectrum of Lyapunov indices.

\begin{figure*}%
\vskip1mm
\includegraphics[width=16.7cm]{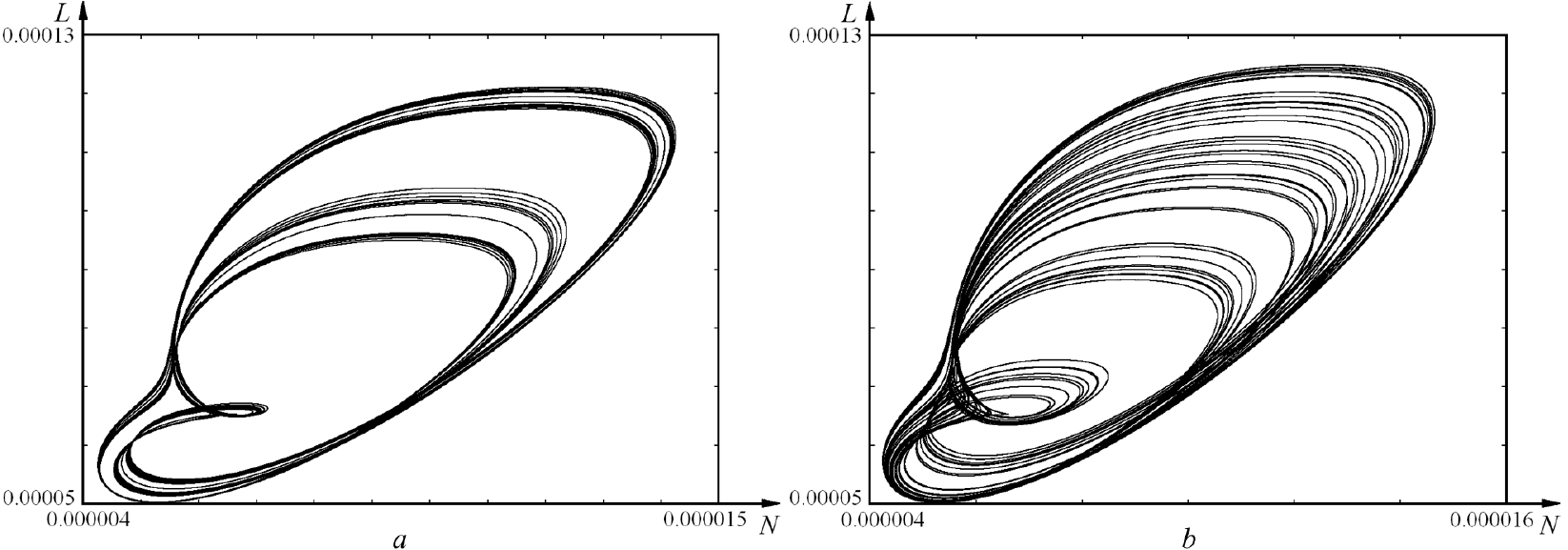}\\[2mm]
\includegraphics[width=16.7cm]{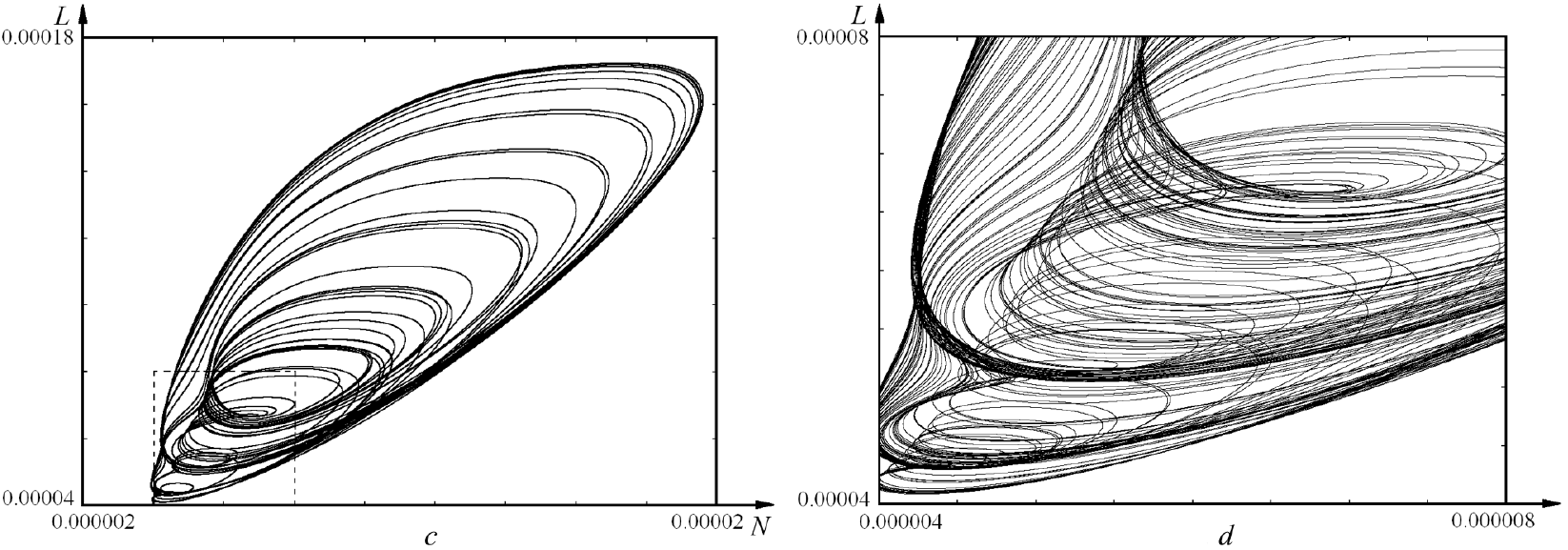}
\vskip-3mm\caption{Projections of the phase portraits of attractors
on the plane $(N,L)$: \textit{a}~-- regular attractor $1\cdot 2^2$
of a quasistable autoperiodic cycle for $G_0 =17.25$, $\lambda _1
=0$; \textit{b}~-- strange attractor $1\cdot 2^x$ for $G_0 =17.2$,
$\lambda _1 =0\mbox{.00004}$; \textit{c}~-- strange attractor
$1\cdot 2^x$ for $G_0 =16.8$, $\lambda _1 =\mbox{.00011}$;
\textit{d}~-- section of the formation of a deterministic chaos in
the mixing funnel (see
Fig.~\ref{fig:5},~\textit{c})}\label{fig:5}\vspace*{1mm}
\end{figure*}

For $G_0 =$ 17.25, the regular attractor $1\cdot 2^2$ of a
quasistable autoperiodic cycle arises.

$\lambda _1 -\lambda _{16} $ is equal to: .00000, --.00004,
--.00006, --.00008, --.00010, --.00054, --.00085, --.00129,
--.00484, --.00562, --.00562, --.00562, --.00980, --.01002,
--.01582, --.02343. $\Lambda  = -.08373$.

For $G_0 =$ 17.2, the strange attractor $1\cdot 2^x$ arises.

$\lambda _1 -\lambda _{16} $ is equal to: .00004, .00000, --.00006,
--.00009, --.00013, --.00053, --.00085, --.00126, --.00483,
--.00556, --.00556, --.00556, --.00972, --.00997, --.01588,
--.02329. $\Lambda  = -.08325$.

For $G_0 =$ 16.8, the strange attractor $1\cdot 2^x$ arises.

$\lambda _1 -\lambda _{16} $ is equal to: .00011, .00000, --.00008,
--.00009, --.00011, --.00057, --.00082, --.00136, --.00480,
--.005700, --.00570, --.00570, --.00937, --.00997, --.01565,
--.02348. $\Lambda  = -.08329$.

For $G_0 =$ 16.5, the strange attractor $1\cdot 2^x$ arises.

$\lambda _1 -\lambda _{16} $ is equal to: .00009, .00000, --.00009,
--.00009, --.00009, --.00057, --.00082, --.00126, --.00475,
--.00560, --.00560, --.00560, --.00903, --.00997, --.01581,
--.02325. $\Lambda  = -.08244$.

For $G_0 =$ 16.3, the strange attractor $1\cdot 2^x$ arises.

$\lambda _1 -\lambda _{16} $ is equal to: .00004, .00000, --.00008,
--.00009, --.00011, --.00055, --.00081, --.00108, --.00462,
--.00505, --.00505, --.00505, --.00917, --.00996, --.01676,
--.02213. $\Lambda  = -.08047$.

In Fig.~\ref{fig:5},~\textit{a--d}, we show the projections of the
phase portraits of attractors in the plane $(N,L)$, which are formed
for the following values of $G_0 $: 17.25; 17,2; 16,8. The larger
the positive senior Lyapunov index $\lambda _1 $, the more unstable
is the system (compare Fig.~\ref{fig:5},~\textit{a--c}).

In Fig.~\ref{fig:5},~\textit{c}, we separate a small rectangular
section, which cover the funnel of mixing of the given strange
attractor, and magnify it (Fig.~\ref{fig:5},~\textit{d}). Let us
calculate the trajectory on a longer time interval. As is seen, the
character of the construction of trajectories of the given strange
attractor is repeated on small and large scales of the projection of
a phase portrait. Every arising curve of the projection of a chaotic
attractor is a source of formation of new curves. Moreover, the
geometric regularity of the construction of trajectories in the
phase space is repeated. The given geometric structure reminds a
two-scale parametric Cantor set.

By using the Lyapunov indices for strange attractors, we determined
the KS-entropy (entropy by Kolmogorov--Sinai) \cite{38}. According
to the Pesin theorem \cite{39}, the KS-entropy $h $corresponds to
the sum of all positive Lyapunov characteristic indices:

The KS-entropy allows one to judge about the rate of loss of the
information about the initial state of the system. The positivity of
the entropy is a criterion of. This gives possibility to
qualitatively estimate the property of local stability of the
attractor.

We determined also the quantity reciprocal to the KS-entropy,
$t_{\min } $. It is the time of mixing in the system, which
indicates how rapidly the initial conditions will be forgotten. For
$t\ll t_{\min } $, the behavior of the system can be predicted with
sufficient accuracy. For $t>t_{\min } $, only its probabilistic
description is possible. The chaotic mode is unpredictable due to
the loss of the memory of initial conditions. The value of $t_{\min
} $ is called the Lyapunov index and characterizes the ``horizon of
predictability'' of a strange attractor.

To classify the geometric structure of strange attractors, we
calculated the value of their fractality. The strange attractors are
fractal sets and have a fractional Hausdorff--Besicovitch dimension.
But its direct calculation is a very laborious problem, which has no
standard algorithm. Therefore, as the quantitative measure of
fractality, we calculated the Lyapunov dimension of attractors by
the Kaplan--Yorke formula \cite{40,41}:
\begin{equation}
\label{eq17} D_{F_r } =m+\frac{\sum\limits_{i=1}^m {\lambda _i }
}{\left| {\lambda _{m+1} } \right|},
\end{equation}
where $m$ is the number of the first Lyapunov indices ordered by
diminution, whose sum is $\sum_{i=1}^m {\lambda _i } \geqslant 0$;
$m+1$ is the number of the first Lyapunov index, whose value
$\lambda _{m+1} <0$.

For the above-presented strange attractors $2^x$, we calculated the
following parameters.

For $G_0 =17.2$: $h = 0.00004$, $t_{\min } =25000$, $D_{F_r }
=2.667$.

For $G_0 =16.8$: $h = 0.00011$, $t_{\min } =9090.9$, $D_{F_r }
=3.375$.

For $G_0 =16.5$: $h = 0.00009$, $t_{\min } =11111.1$, $D_{F_r }
=3.000$.

For $G_0 =16.3$: $h = 0.00004$, $t_{\min } =25000$, $D_{F_r }
=2.500$.

These results allow us to judge about the difference of the
geometric structures of the given strange attractors. The largest
value of KS-entropy is obtained for $G_0 =16.8$. This indicates the
highest chaotic mixing of trajectories in a funnel. Respectively,
the mixing time, after which the unpredictable chaos arises, will be
minimal. The Lyapunov dimension for the given attractor, which
characterizes quantitatively the fractality, is maximal in this
case. This is confirmed by Fig.~\ref{fig:5},~\textit{c},~\textit{d}.
By calculating successively the different strange attractors, we can
determine a certain regularity in the hierarchy of their chaotic
behavior. The geometric shape of attractors of the system varies
correspondingly to the change of these indices. Thus, glycolysis in
a cell is adapted at the variation of the amount of external glucose
($G_0 )$ to the varying conditions of the external environment, by
preserving its functionality in this case.

\section{Conclusions}

With the help of the mathematical model, we have studied the
metabolic process of glycolysis arising as a product of the
biochemical evolution in protobionts. It is shown that glycolysis
can be distinguished as a united self-regulating complex of the
metabolic network of a cell. The conditions of its self-organization
in the cyclic mode are determined. The bifurcations of doubling of a
cycle according to the Feigenbaum scenario are discovered, and it is
found that the intermittence results finally in the appearance of
aperiodic modes of strange attractors. This means that the intensity
of the metabolic process of glycolysis is adapted to the varying
conditions of the medium. The fractal nature of the obtained cascade
of bifurcations is demonstrated. The strange attractors, which are
formed due to the formation of a mixing funnel, are found. The
complete spectra of Lyapunov indices and the divergences for various
modes are calculated. For the strange attractors, the KS-entropies,
``horizons of predictability,'' and the Lyapunov dimensions of
attractors are determined. The obtained results allow one to study
the structural-functional connections of the metabolic process of
glycolysis and their influence on the cyclicity of metabolic
oscillations in a cell, as well as to clarify the physical laws of
the self-organization in it.

\vskip3mm \textit{The work is supported by the project
No.\,0112U000056 of the National Academy of Sci\-en\-ses of
Ukraine.}

\rezume{%
В.Й.\,Грицай}{САМООРГАНІЗАЦІЯ І ФРАКТАЛЬНІСТЬ\\ В МЕТАБОЛІЧНОМУ
ПРОЦЕСІ ГЛІКОЛІЗУ} {В роботі за допомогою математичної моделі
досліджується метаболічний процес гліколізу. Загальна схема
гліколізу розглядається як закономірний результат біохімічної
еволюції. Використовуючи теорію дисипативних структур, проведено
пошук умов самоорганізації даного процесу. Знайдено автокаталітичні
процеси, завдяки яким зберігається циклічність в динаміці його
протікання. Досліджено умови порушення синхронізації процесу,
збільшення кратності циклічності та виникнення хаотичних режимів.
Отримані фазопараметричні діаграми каскаду біфуркацій, які
відзеркалюють перехід до хаотичних режимів відповідно сценарію
Фейгенбаума та перемежаємості. Знайдено дивні аттрактори, що
виникають внаслідок воронки. Побудовані їх аттрактори. Розраховані
повні спектри показників Ляпунова і дивергенцій для знайдених
режимів. Розраховані КС-ентропії, горизонти передбачуваності та
ляпуновскі розмірності дивних аттракторів. Зроблено висновки про
структурно-функціональні зв'язки гліколизу, їх  впливу на стйкість
метаболічного процесу клітини.}

\rezume{%
В.И.\,Грицай\vspace*{1mm}}{САМООРГАНИЗАЦИЯ И ФРАКТАЛЬНОСТЬ\\ В
МЕТАБОЛИЧЕСКОМ ПРОЦЕССЕ ГЛИКОЛИЗА\vspace*{1.0mm}} {\rule{0pt}{12pt}В
работе при помощи математической модели исследуется метаболический
процесс гликолиза. Общая схема гликолиза рассматривается как
закономерный результат биохимической эволюции. Используя теорию
диссипативных структур, проведен поиск условий самоорганизации
данного процесса. Найдены автокаталитические процессы, вследствие
которых сохраняется цикличность в динамике его протекания.
Исследованы условия нарушения синхронизации процесса, увеличения
кратности цикличности и возникновения хаотических режимов. Получены
фазопараметрические диаграммы каскада бифуркаций, отражающие переход
к хаотическим режимам согласно сценарию Фейгенбаума и
перемежаемости. Найдены странные аттракторы, образуемые в результате
воронки. Построены их аттракторы. Расчитаны полные спектры
показателей Ляпунова и дивергенций для найденных режимов. Рассчитаны
КС-энтропии, горизонты предсказуемости и ляпуновские размерности
странных аттракторов. Сделаны выводы о структурно-функциональных
связях гликолиза, их влияния на устойчивость метаболического
процесса клетки. }

\end{document}